\documentclass[10pt,conference]{IEEEtran}
\IEEEoverridecommandlockouts
\usepackage{cite}
\usepackage{amsmath,amssymb,amsfonts}
\usepackage{algorithmic}
\usepackage{graphicx}
\usepackage{textcomp}
\usepackage{xcolor}
\def\BibTeX{{\rm B\kern-.05em{\sc i\kern-.025em b}\kern-.08em
    T\kern-.1667em\lower.7ex\hbox{E}\kern-.125emX}}

\usepackage{graphicx} % Required for inserting images
\usepackage{hhline}
\usepackage{orcidlink}

\usepackage{commands}

\begin{document}

\title{Which Alert Removals are Beneficial?}
%\author{Idan Amit, \orcidlink{0009-0007-7881-9090}}

\author{\IEEEauthorblockN{Idan Amit \orcidlink{0009-0007-7881-9090}}
}

\maketitle

\begin{abstract}
\emph{Context}
Static analysis captures software engineering knowledge and alerts on possibly problematic patterns.
Previous work showed that they indeed have predictive power for various problems.
However, the impact of removing the alerts is unclear.
\emph{Aim}
We would like to evaluate the impact of alert removals on code complexity and the tendency to bugs.
\emph{Method}
We evaluate the impact of removing alerts using three complementary methods.
1. We conducted a randomized controlled trial and  built a dataset of 521 manual alert-removing interventions
2. We profiled intervention-like events using labeling functions.
We applied these labeling functions to code commits, found intervention-like natural events, and used them to analyze the impact on the tendency to bugs.
3. We built a dataset of 8,245 alert removals, more than 15 times larger than our dataset of manual interventions. 
We applied supervised learning to the alert removals, aiming to predict their impact on the tendency to bugs. 
\emph{Results}
We identified complexity-reducing interventions that reduce the probability of future bugs. Such interventions are relevant to 33\% of Python files and might reduce the tendency to bugs by 5.5 percentage points.
\emph{Conclusions}
We presented methods to evaluate the impact of interventions.
The methods can identify a large number of natural interventions that are highly needed in causality research in many domains.
\end{abstract}

%\begin{IEEEkeywords}
%Machine learning, Software Engineering, Text analysis
%\end{IEEEkeywords}

% Another online causality book
% https://matheusfacure.github.io/python-causality-handbook/landing-page.html

\section{Introduction}

Static analyzers alert on code patterns that might be problematic, such as a too-many-branches or a simplifiable-if.
The alerts are based on domain knowledge, which shows agreement with the presence of bugs \cite{trautsch2020longitudinal}, code review comments \cite{zampetti2022using}, and low grades by a teacher \cite{edwards2017investigating}. 
On the other hand, the usage of static analyzers involves problems such as false positives \cite{ 10.1145/1368088.1368135} %, 5770619, kremenek2003z}
, which seems to limit their use and reduce their efficiency \cite{6606613, lenarduzzi2019does}.

Intervention experiments, conducted in a randomized controlled trial design, are the gold standard method for causality research \cite{jadad1998randomised, stolberg2004randomized,stanley2007design, bhide2018simplified}.
A researcher randomly divides objects of interest into an intervention and a control group. 
Interventions are performed on the intervention group.
The impact is evaluated by comparing the objects in the intervention group before and after the intervention and by comparing the intervention group with the control group. 
The effort required for such experiments might be high, leading to a small dataset.
The size leads to threats of sensitivity to noise and a lack of representation of areas of interest.
Optimally, we would like to maximize the amount of information that can be generated from such an effort, i.e., enjoying the benefits of clean analysis and high observational volume.
We do that by identifying natural events that perform interventions and use them to extend our scale of analysis. 

Causality research can find out which alerts are beneficial in a given context.
Recommendations based on such findings will allow developers to invest effort effectively.

We used three different methods to find which alerts removal is beneficial.
We start with a randomized controlled trial of manual intervention.
This method provides a limited number of clean samples.

The second method profiles intervention-like events in live projects.
This method provides more samples, yet some of them might not be clean interventions (e.g., they might add some logic other than just refactoring to fix the alert).
The method was implemented using the following steps.
We built alert removal labeling functions based on domain knowledge and manual interventions.
We validated the labeling functions using manually labeled samples.
We later ran the labeling functions on a large dataset of open-source commits, identifying intervention-like events in nature.
Since these commits occurred in live projects, we can examine the context before and after the intervention and identify changes in the tendency to bugs.

In the third method, we use the entire large alert removals dataset and we use supervised learning to identify which alert removals are beneficial.
This method might identify beneficial removals that change the code logic.
On one hand, it might recommend on none-applicable removals (e.g., removing an alert by deleting the entire function in which the alert resides)
On the other hand, it is aware of the removal context and can identify beneficial interventions that we did not consider before.

Our main contributions are:
\begin{itemize}
  \item Introducing methods for identifying natural events as interventions, enabling causal analysis on large datasets. In this research, it produced more than 15 times more removals than manual work, and the potential is higher. This method is not specific to software engineering and can be beneficial in many fields. 
  \item Providing developers with simple causality-verified recommendations to reduce the tendency to bugs. 
  \item Introducing manual interventions as a method for researchers to investigate causal influence in static analysis and building a dataset of such interventions.
  \item Handling causal discovery and policy learning by supervised learning on a benchmark dataset of interventions, all important challenges in causality \cite{cinelli2025challengesstatisticsdozenchallenges}.
\end{itemize}

\section{Related Work}

\subsection{Static Analysis}

Static analysis \cite{nagappan05, 4602670}
is a source code based analysis, as opposed to dynamic analysis in which the code is being executed.
Static analysis detects predetermined patterns that might be harmful based on domain knowledge.
Static analysis allows us to automatically analyze large amounts of code and identify patterns that developers might neglect or not be aware of their importance.
Therefore, static analysis tools are useful to predict bugs \cite{ayewah2008using}, vulnerabilities \cite{chess2004static}, and technical debt \cite{zazworka2014comparing}.
However, alert instances are not necessarily problems, but warnings.
There are no requirements for alert definition, and developers implementing static analyzers choose their alerts.
While they capture knowledge, their value is not always clear, a gap that we would like to fill.
In fact, the value of the static analysis was criticized \cite{ 10.1145/1368088.1368135, 5770619, kremenek2003z}.
Sometimes developers do not: tend to use static analysis tools \cite{6606613}, choose refactoring targets based on static analysis, or consider alerts in pull request approval \cite{lenarduzzi2019does}.

Our work investigated the benefit of alerts on code metrics and the tendency to bugs. 
It resembles the work of Basili et al.\ \cite{544352} who validated the K\&C metrics \cite{Chidamber:1994:MSO:630808.631131} as predictors of fault-prone classes.
Couto et al.\ \cite{couto2014predicting} used Granger causality tests, showing additional predictive power in time series \cite{granger1969investigating}, to predict defects.
Trautsch et al.\  investigated the evolution of PMD static analyzer alerts and code quality, operationalized as defect density \cite{trautsch2020longitudinal}.
Edwards et al. investigated the relationship between students' grades and alert categories of CheckStyle and PMD \cite{edwards2017investigating}.
Zampetti et al.\ evaluated CheckStyle alerts according to their agreement with code review comments \cite{zampetti2022using}.
Note that in both code review and grading, there is a person who recommends a change, as in the causal use case.

\subsection{Causality}

Causality research is a method to investigate the impact of actions. 
Randomized controlled trials  
are considered the gold standard for determining the impact of an action \cite{jadad1998randomised, stolberg2004randomized,stanley2007design, bhide2018simplified}.
However, the effort needed per intervention tends to limit the number of interventions, threatening the validity of the results by high sensitivity to outliers or even violating the assumption of equivalence of the control and intervention groups \cite{deaton2018understanding, doi:10.1056/NEJMra1614394}. 

Observational data analysis is used to leverage larger datasets.
Structural Equation Modelling \cite{wright1921correlation} %, haavelmo1944probability, chernozhukov2024applied}
, Rubin Causal Model \cite{sekhon2008neyman}, and Bayesian Networks \cite{pearl2009causality} are used to predict the causal impact using observational data.
However, different causal models might lead to the same observational data, and therefore observational data alone cannot distinguish them \cite{pearl2009causality, cinelli2025challengesstatisticsdozenchallenges}. 

Twin experiments are a special case of comparison of pairs of close cases in which there was an intervention in only one of the members.
Since the members of the pairs are close and differ in the intervention, they allow comparing the impact. 
This family contains twins experiments \cite{vandenberg1966contributions, Amit2024Motivation}, matching \cite{heckman1997matching},  natural experiments \cite{dinardo2018natural}, regression-discontinuity analysis \cite{thistlethwaite1960regression}, and diff in diff \cite{card1993minimum}. %, bertrand2004much}.

Temporal analysis is another approach to leverage observational data, generated over time.
Granger \cite{granger1969investigating} models time series to predict future events.
Co-change analysis predicts the change in the outcome given a change in features, using pairs of the same object in two consecutive periods \cite{Amit2021CCP, amit2021follow, AMIT2025Survey}.

\subsection{Intervention Experiments in Software Engineering}

Intervention experiments are common in experimental sciences (e.g., when investigating the impact of a new drug).
Rochimah et al. intervened and fixed code smells and evaluated their impact on  modularity, analysability, reusability, and testability metrics \cite{11250643}.
Shepperd et al. used interventions to assess the anchoring effect of developers \cite{10.1145/3167132.3167293}.
Nunes et al.\ built a dataset of 693 changes in open-source Java projects, including method extraction, manually verified to be refactoring \cite{nunesmarv}.
Fontana et al. recently refactored about a thousand  architectural smells in Java projects \cite{fontana26impact}. Their results agree with ours and indicate that removing smells improves code quality in general, and specifically reduces the sum of McCabe complexity \cite{mccabe76}.
\hide{
The Minnesota-Linux story is an incident of code modification for research \cite{Chin2021Minnesota}.
The Linux kernel was modified to see if one could inject security-problematic code.
Obviously, developers ``do not appreciate being experimented on'' \cite{feitelson2021we}.
To avoid that, we described our manual intervention experiment in our repository, linked to it, explained it, and asked the owners to merge.
}

\subsection{Labeling Functions}

A labeling function \cite{Blum:1998:CLU:279943.279962} is a validated computable weak classifier, a heuristic that can be applied computationally to a dataset and get predictions that are better than a guess.

The concept of weak learnability \cite{kearns1988learning}, learning slightly better than a guess, was suggested as a way to relax the high requirements of PAC learning \cite{valiant1984theory}.
Surprisingly, it was shown that the concepts are equivalent, and that weak learners can be boosted to regular PAC learners \cite{schapire1990strength}.
Since then, boosting has become an important learning method \cite{schapire1990strength, FREUND1997119}.
Weak classifiers were also found to be powerful in coping with the lack of labeled data.
They were part of both theoretical and practical work such as co-training \cite{Blum:1998:CLU:279943.279962} and weakly supervised learning \cite{Amit2017Archimedes, Amit2024Motivation}.

\hide{
An example of a labeling function is ``Developers that work 9 to 5 have less motivation than those that work in more diverse hours'' \cite{Amit2024Motivation}.
This rule is not perfectly accurate, yet it encapsulates knowledge that improves our prediction.
Given a single labeling function, one can use it as a proxy for the concept.
For example, retention in a project is a labeling function that can be used to investigate the concept of motivation \cite{Amit2024Motivation}.
}
\hide{
The flexibility of labeling functions allows using them in various settings.
Given a few labeling functions, one can evaluate them with respect to each other \cite{Amit2024Motivation}.
The functions can be further improved using techniques such as co-training \cite{Blum:1998:CLU:279943.279962}.
They can be used in active learning algorithms to find informative samples to label \cite{Amit2017Archimedes}. 
}

\section{Method}

\subsection{Manual Intervention Methodology}

Pylint has many alerts, and our intervention capacity is limited.
Therefore, we had to choose specific types of alert for the experiment.
We avoided too rare alerts since their analysis will be sensitive to noise, and developers are less likely to encounter them.
We aimed to avoid alerts that indicate direct errors (e.g., duplicate-argument-name, function-redefined) since they indeed should be fixed for basic correctness. 
We also avoided alerts that might result from diverse reasons and require very different interventions, such as `fixme'. 

All Pylint alerts were implemented since they were considered beneficial in the first place.
However, we deliberately choose a few alerts whose benefit is expected to be very low, to serve as negative controls \cite{shi2020selective}.
For example, superfluous-parens, unneeded extra parentheses, are common in return statements and conditions \cite{zampetti2022using,amit2021follow}.
The extra parentheses add two characters and slow down reading.
Baron et el. showed that there is no significant difference in the time to comprehend different De-Morgan representations of the same formula \cite{10.1145/3661167.3661180}.
Hence, the time difference to comprehend a formula with the smaller difference of extra parentheses is likely to be small too.
The negative controls allow distinguishing the direct influence of the intervention and the indirect one (e.g., drawing attention to the modified file, which leads to another improvement).

Given the alerts, we needed to select the candidates for manual intervention.
We wrote a utility that runs Pylint on all Python files (excluding tests) in a repository.
The utility randomly splits the files into control and intervention groups.
For each file, a single alert type is chosen for intervention, which must not appear more than twice in the file, to keep the interventions small.
We fixed all the intervention candidates in a small, focused refactoring and submitted a pull request to the owners.
The projects that we used are up-to-date large software repositories, whose selection is detailed in previous work \cite{Amit2021CCP, amit2021end, Amit2019Refactoring}.
\hide{
Intervention protocol:
\begin{enumerate}
  \item Pick a Python GitHub open-source project. We created a list of candidate projects from previous work \cite{Amit2021CCP, amit2021end, Amit2019Refactoring}.
  \item Open an issue declaring the experiment, link to the experiment repository, and ask for approval to intervene.
  \item Fork the project on GitHub and get familiar with it. 
  \item Run our utility to get intervention candidates.
  \item Fix all candidate alerts and verify that by rerunning Pylint. Please make sure that your change is minimal. Avoid improving the file in other ways.
  \item Create a pull request, submit the pull request, modify it according to the feedback, and ask the owner to merge.
\end{enumerate}
}

\subsection{Intervention-Like Events Methodology}

To scale our analysis, we needed an effective way to identify intervention-like events in nature.
We used frozen versions of our repositories, taken in February 2022 and August 2022 \cite{amit2021end}.
Of the 449,512 Python files in February 2022, only 103,505 (23\%) were changed by August, allowing us to filter out files in which the alerts remained unchanged.
In 54,855 of the February files that were changed, we found 109,895 alerts of interest.
Only 5,870 (5\%) of the alerts that appeared in February, no longer appeared in August.
In these cases, we were assured that the file history contains a removal of the alert.
We ran Pylint on the commits timeline of these files, looking for all changes in commits, finding 8,245 alert changes.
This is more than 15 times more than the 521 manual interventions that we performed (Table \ref{tab:alerts}).
An average commit duration is 83 minutes \cite{Amit2020Effort}, therefore building a dataset of 8,245 alert removals could take 475 days.

We evaluated the tendency to bugs using the Corrective Commit Probability (CCP) process metric, which uses the commit message to decide whether the commit was a bug fix \cite{Amit2021CCP}.
To estimate the CCP metric, we required at least five commits before and after the change, getting 3,409 such alert removals.
For comparison, 271 of our manual interventions were merged, not necessarily having enough commits before and after them.
Therefore, this methodology produced at least 12 times more commits.
More than that, we applied the method using only two specific dates, hence running automatic processes, with minimal human effort, can potentially lead to many more cases.

Not all alert removals are due to refactoring.
For example, 6\% of these commits were only code deletion.
To automatically filter out commits that did not remove the alert due to refactoring,
we used domain-knowledge, and familiarity with alert removal that was gained while building the manual interventions dataset.

Consider the intervention "extract method", which is called the “Swiss army knife” of refactoring \cite{10.5555/2555523.2555539, silva2016we, hora2020characteristics, 10381511}.
Possible labeling functions for it are: reference to method extraction in the commit message, reduction in the maximal McCabe complexity, and introduction of new functions. 
We used all the above directions in the analysis.

The labeling functions can run automatically on all commits. 
We composed them from smaller clauses, not predicting a refactor on its own but being used in a larger function.
For example, we created a clause that checks that the commit also adds lines.
Some identified directly a refactoring-related property (e.g., adding a new function to the file, as done when extracting a method).
We sampled and labeled commits to evaluate the performance of the labeling functions.
We list the functions and their performance in Section \ref{sect:lb-results}. 

\subsection{Supervised Learning Modeling of Impact}

So far, we have evaluated hypotheses regarding reducing the tendency to bugs that were grounded in prior work.
Now we move to exploring the dataset to predict a reduction in the tendency to bugs.
We do that by applying supervised learning to model changes \cite{AMIT2025Survey}.
Once we have the interventions and their context, we can represent a concept of ``bug tendency reduction'' and apply supervised learning classifiers on the context variables to identify beneficial actions, important challenges of causal discovery and policy learning \cite{cinelli2025challengesstatisticsdozenchallenges}.

Since we reduced the problem into a classification problem, we can use existing tools for it.
We used the scikit-learn package for classification algorithms \cite{scikit-learn}.
We used low-capacity small models such as decision trees and logistic regression to obtain simple interpretable models that are also rather robust to overfitting \cite{arpit2017closer}.
We also used models of medium capacity: random forests, k-nearest neighbors, AdaBoost, gradient boosting, support vector machines, stochastic gradient descent, and neural networks to build models of better representation ability and performance.

This method uses more samples than the two previous ones, allowing us to cover more cases and higher statistical confidence.
On the other hand, we analyze samples where the removal is not a clean refactoring intervention.
Hence, though we find that just deleting code is beneficial, this is not a generally applicable recommendation.

\section{Results}

\subsection{Manual Interventions and Code Metrics}

Now we analyze the impact of our manual interventions on code metrics.
The code contains all the logic and implementation details.
The intervention is the only change, leading to an experiment with reduced noise.
We used the classical McCabe complexity \cite{mccabe76}, counting branching statements.
The number of paths in a function might be exponential in its McCabe complexity (e.g., in a linear if-statements sequence).
We also used lines of code (LOC), which most other code metrics are correlated with \cite{gil17}.
We present the mean change in each metric per alert type.
Halstead metrics \cite{halstead1977elements} are provided in our repository. %\cite{Amit2025Pylint}.

Table \ref{tab:alerts} presents the interventions and their impact on code metrics.
It provides the alert and its description.
`Repos.' is the number of repositories containing the alerts.
`Alerts' is the number of cases considered, including cases where fixing the alert is not beneficial.
`Merged' are modified alerts that are merged into the repository, which can be used for process metrics analysis.
The other columns are the average of the metric value after the intervention minus the metric value before the intervention.
`McCabe Max' is the reduction in the maximal McCabe metric for a function (the one with the original alert).
`McCabe Sum' is the sum difference of McCabe metrics for all functions in the file.
`LOC' is lines of code difference.

The McCabe metric \cite{mccabe76}, introduced in 1976, is a fundamental code metric.
It counts branches in functions as a measure of complexity.
The number of paths in the function, important for testing all options, might be exponential in the McCabe metric, which affects testing methods \cite{watson1996structured}.
The McCabe metric is an important predictor of defects \cite{menzies2007data}, maintenance ability \cite{1702603}, and the ability of developers to detect defects \cite{10.5555/800091.802959}.
That makes the McCabe metric an important stepping stone, shedding light on a mechanism in which an intervention makes a direct change on the code, captured in a reduction in the McCabe metric and translated into a process metric of lower tendency to bugs.

%The table lists 17 alert types in various repositories. 
The main alerts of interest were considered more than 30 times, a common lower bound for statistical analysis.
We considered all candidate alerts in a project, to avoid selection bias.
The high difference in the number of alerts is due to the prevalence of the alerts.

An interesting result is the reduction in `McCabe Max' in the alerts that indicate function complexity.
The highest average reduction is 13.6 for too-many-statements (all its interventions reduce at least 5 points) and the lowest is 5.6 for too-many-return-statements, with too-many-branches and too-many-nested-blocks in between.
Note that the number of code paths can be exponential in McCabe hence they are between $2^{5.6}=48$ and $2^{13.6}=12,416$ saved paths.
This is very important for testing and using the code.
Hence, our experiment shows that fixing these alerts has a high impact on the code.
While method extraction is a common way to fix these alerts, in too-many-return-statements one can refactor to store the result in a variable and return it at the end of the function, keeping the McCabe metric unchanged.

If we merely encapsulate a part of a function in a new function, the sum of the McCabe metrics should not change.
It might increase in the case of integrating the new function which requires new ifs (increase in 5 to a maximum of 9 points in 10\% of too-many-branches).
This is common when using multiple conditional returns, forcing the parent function to check if it should return with new if statements.
If the extracted function is reused, then `McCabe Sum' is also reduced (32\% of too-many-branches).

simplifiable-if-expression and simplifiable-if-statement remove a single unneeded if, reducing the McCabe metric by one, making it convenient for process metrics analysis.

Note that the LOC differs in small absolute values in all cases.
That, along with personal observation regarding the required time, indicated that the investment in fixing the alerts is not high.

\begin {table*}[t]\centering
\caption{ \label{tab:alerts} Interventions Code Metrics Difference }
\begin{tabular}{ | p{40mm}|  p{70mm}| p{8mm}| p{8mm}| p{8mm}| p{9mm}| p{9mm}| p{8mm}  | }
\hline
Alert & Description & Repos. & Alerts & Merged & McCabe Max & McCabe Sum & LOC \\
\hline
broad-exception-caught & Catch Exception, might hide unintended exceptions & 14  & 74  & 45  & 0.4 & 0.1 & 0.9\\ \hline
comparison-of-constants & A comparison whose result is known, since the sides are constants. Usually a bug & 2  & 2  &  &  & 0.0 & 0.0\\ \hline
line-too-long & Might hide important code & 17  & 161  & 93  &  & 0.0 & 1.6\\ \hline
pointless-statement & Usually unintended & 3  & 5  &  &  & 0.0 & -1.0\\ \hline
simplifiable-if-expression & True if cond else False, equivalent to cond & 4  & 17  & 16  & -1.0 & -1.1 & 0.0\\ \hline
simplifiable-if-statement & As the expression yet with a full if & 3  & 8  & 6  & -1.0 & -1.0 & -3.0\\ \hline
superfluous-parens & Unneeded extra parenthesis, common in return statements and conditions \cite{zampetti2022using,amit2021follow}. & 9  & 33  & 38  &  & 0.0 & 0.0\\ \hline
too-many-boolean-expressions & An if with many terms & 2  & 3  & 1  & -1.0 & -3.0 & 1.5\\ \hline
too-many-branches & A function/method with too many if statements & 15  & 73  & 32  & -8.3 & -0.4 & 2.7\\ \hline
too-many-lines & A file with too many lines & 2  & 6  &  &  &  & \\ \hline
too-many-nested-blocks & A function/method with too high nesting level \cite{lenarduzzi2020sonarqube, amit2021follow, zhang2018automated}  & 4  & 10  & 4  & -10.1 & 0.4 & 3.6\\ \hline
too-many-return-statements & A function/method with too many return statements & 6  & 31  & 7  & -5.6 & -2.6 & 0.1\\ \hline
too-many-statements & A function/method with too many lines & 8  & 20  & 7  & -13.6 & 2.6 & 6.1\\ \hline
try-except-raise & Might be a useless error handling & 1  & 1  &  &  &  & \\ \hline
unnecessary-pass & Sometimes a useless statement, many time an empty class & 10  & 63  & 18  &  & -0.2 & -1.2\\ \hline
using-constant-test & An if whose result is known. Usually a bug & 1  & 2  &  &  &  & \\ \hline
wildcard-import & Import *. Leads to unclear source and possible future collisions \cite{lenarduzzi2020sonarqube, kery2016examining, bestPractices, amit2021follow} & 7  & 12  & 4  &  & 0.0 & 0.2\\ \hline  
\end{tabular}
\end{table*}

\subsection{Intervention-like Events and Tendency to Bugs}\label{sect:lb-results}

\subsubsection{Labeling Functions Validation}

An intervention should remove the alert in a refactor, keeping the previous logic, removing the alert, and doing only that (e.g., avoiding tangling commits \cite{herzig2013impact, herbold2022fine}).
Commit messages are used to describe the essence of the change \cite{choshen2021comsum}, hence an indication of refactoring in alert removal commit provides external evidence of its nature.
While in many projects the refactoring rate is about 10\%,  32\% of our alert removal commits are refactoring.
This is a high increase, yet for many of the removals we lack this additional evidence of refactoring. 
In 6\% of the commits, the modification did not add any lines, and hence the removal is not due to refactoring but due to deletion.
14\% of the commits are large and have more than 300 modified lines, and hence it is probably not a focused one.
Minor alerts, of low importance and effort, are often not removed in a dedicated commit but as part of a massive change (e.g., too-many-boolean-expressions in 37\%, wildcard-import in 31\%, superfluous-parens in 21\%).
Single-line alerts also tend to be part of commits that modify more lines (e.g., superfluous-parens and unnecessary-semicolon in 90\%, simplifiable-if-expression in 87\%).

To identify commits that are likely to be interventions, we use labeling functions, listed in Table \ref{tab:lf-pred-performance}.
The upper part lists the clauses that are used by the labeling functions in the lower part.
The clauses predict various properties of the change, and the labeling functions predict that a change is a refactor.
The predictive performance of the labeling functions is based on manual labeling of alert-removal commits.
The functions are applied to commits in which a function complexity alert is removed ('too-many-branches', 'too-many-nested-blocks', 'too-many-return-statements', 'too-many-statements')

`Recall' is the probability of an alert removal identification by the function out of the alert removals, and `Refactor precision' is the commit that indeed refactored the code.
We measure the performance of each clause in the context of the labeling function in which they are used (e.g., McCabe reduction and code addition).

\begin {table*}[t]\centering
\caption{ \label{tab:lf-pred-performance} Labeling Functions predictive performance }
%\begin{tabular}{|l| p{70mm}|p{10mm}|p{10mm}|p{10mm}|} \hline
\begin{tabular}{|p{40mm}| p{70mm}|p{25mm}|p{25mm}|} \hline
Alert & Description & Recall  &  Refactor Precision \\
\hline
New line in commit &  & 100\%  &   \\ \hline
New function addition &  & 80\%   &  88\% \\ \hline
Massive change & At least 300 lines changed. Used in negation. & 0\%  &   \\ \hline
Mostly delete & Deleted lines are no more than 3 times the added lines. Used in negation. & 0\% (suitable)  &   \\ \hline
Max McCabe complexity reduction & The maximum McCabe of the functions in the file is lower & 85\% (suitable), 15\% (reduced)    &  71\% (suitable), 66\% (reduced) \\ \hhline{|=|=|=|=|}
Reduced McCabe refactor  & Max McCabe complexity reduction and new lines in commit (computable from the diff) & 64\%   &  60\% \\  \hline
Suitable McCabe refactor  & Max McCabe complexity reduction, new lines in commit (computable from the diff), not mostly delete, not massive change & 48\%   &  55\% \\  \hline
Added function refactor  & A new function was added in the commit & 39\%   &  80\% \\  \hline
\end{tabular}
\end{table*}

First, note that the clauses' predictive performance is almost always perfect (100\% or 0\% when used in negation).
This is because they are a simple computation on the commit.
Even with new-function-addition, with 80\% precision, the false positives are bizarre cases such as a function rename or turning an existing function into an asynchronous one.

The recall of the labeling functions is high, which shows that we do not capture only a few of the cases.
The precision ranges from good to very high.
Usually, the more specific clauses restrict a function scope, the higher the precision and the lower the recall.
However, the reduced McCabe has a bit higher precision than the suitable McCabe, since it identifies a refactoring in which a function outside the file is used.

\subsubsection{Interventions Impact on Tendency to Bugs}

Given the validated labeling functions, we evaluated the impact on the tendency to bugs in various contexts.
We present in Table \ref{tab:ccp-diff} some beneficial interventions and their impact on reducing the tendency to bugs.
It includes simple recommendations for interventions that solve common severe problems that developers can easily apply.
The `Context' is when the removal occurs, `Alert' is the removed alert, and `CCP difference' is the average difference between the CCP after the removal and the CCP before.
`Samples' is the number of cases and `Std. Error' is the standard error $\frac{std.}{\sqrt{samples}}$.
In the remainder of this section, we provide context and discuss the impact of interventions.

\begin {table*}[t]\centering
\caption{ \label{tab:ccp-diff} Controls and Bug Tendency Reducing Interventions }
\begin{tabular}{|l| l|c | c| c|c|} \hline
Context & Alert & CCP difference  (percentage points) & Samples & Std. Error & McCabe\\
\hline
-  & Refactoring commit  \cite{Amit2019Refactoring} & 0.2   & 176,309 & &\\ \hline
Self Admitted Tech Debt removal & TODO \cite{Amit2019Refactoring} & -0.3    & 256 & 0.8 &\\ \hline
Single line & superfluous-parens & 0.1     & 34 & 2 &\\ \hline
New function & too-many-nested-blocks & -5.5       & 35 & 5.0 & -10.8 \\ \hline
New function & too-many-branches & -4.1   & 162 & 1.9 & -9.1 \\ \hline
Suitable McCabe refactor & too-many-return-statements & -5.0 & 80 & 2.0 & -3.8 \\ \hline
\end{tabular}
\end{table*}

We measured the tendency to bugs using Corrective Commit Probability (CCP), based on the ratio of bug fix commits \cite{Amit2021CCP}.
CCP is a process metric, not based on the code itself, avoiding direct dependence on it and circular relations between the alert and the metric.
We identified 3,409 alert removal cases, which had at least 5 commits both before and after them, to reduce sensitivity to noise.

As a control group, we consider regular development without intervention.
The average file CCP increases by 0.05 percentage points in 3 months after a refactoring, equivalent to 0.2 points a year \cite{Amit2019Refactoring}.
Per repository, the average CCP increases by 0.6 points compared to the year before \cite{Amit2021CCP}.
We also observed our single-line negative controls.
The behavior of superfluous-parens is, as expected, close to zero, yet better than the regular commit control.
We wanted to use line-too-long as a negative control too.
However, since many times it is removed as part of a larger modification, and not all interventions have enough commits in their context, we had only three such cases.

Extracting code into a new function is a common way to reduce complexity, as demonstrated in the reduction in the McCabe metric.
Our main result is that when adding a function while removing too-many-branches the average CCP reduces in 4.1 percentage points, and when removing too-many-nested-blocks it reduces by 5.5 percentage points.
Note that within the alert, the reduction in the McCabe metric is not ordered like the reduction in bugs, indicating the existence of more influencing factors.
To our knowledge, we are the first to show that an action \emph{causes} a reduction in the tendency to bugs.
However, Lenarduzzi et al. also found that the SonarCube rule equivalent of  too-many-nested-blocks is beneficial in defect \emph{prediction}, supporting our result \cite{lenarduzzi2020sonarqube}.
As a benchmark, consider removing Self Admitted Tech Debt \cite{potdar2014exploratory} (e.g., `TODO'),  being the most effective refactoring explored by Amit and Feitelson \cite{Amit2019Refactoring}.
Removing `TODO' reduces CCP 0.3 percentage points, and the impact of removing too-many-nested-blocks by adding a new function is more than 18 times better.

It is interesting to note that adding a new function when removing too-many-return-statements has a positive CCP difference of 2 points and too-many-statements has a 4.2 point diff.
This might be because branches (and the McCabe metric) better represent complexity than statements (and the LOC metric) and returns, and possibly the original state is not necessarily problematic when the alert is regarding the last two. 
Alert removals that added new functions without reducing the McCabe complexity could shed light on this case, yet we had too few such cases.
However, in `Suitable McCabe refactor', where we require McCabe reduction,  too-many-return-statements has a difference of -5 points, and too-many-statements has a positive difference of 2.3.

We have two alerts that remove a single local unneeded if (e.g., using ``True if x else False'' instead of ``bool(x)'' or even just x if it is binary), allowing the examination of a simple alert and intervention.
The CCP difference for the single-line  simplifiable-if-expression is 0.9, and for simplifiable-if-statement is 4.3.
These alerts are rather similar, and though the simplifiable-if-statement is a bit more complex, the gap between them might indicate the effect of noise in the data.

We would like to see that the CCP difference is not due to regression to the mean \cite{barnett2005regression}.
We divided the CCP levels into low (the lowest 25\%, CPP $<$ 9\%), high (the highest 25\%, CCP $>$ 39\%), and medium for the middle.
The CCP difference is heavily influenced by the original level with 15.9 for low, 3.1 for medium, and -16.3 for high.
Such monotonicity tends to happen even per alert (e.g., low impact line-too-long, single line simplifiable-if-expression, impactful too-many-branches).
Note that for fewer than 11 commits, only 0 fixes will lead to less than 0.09 CCP.
Hence, in many low CCP cases results cannot be improved, and no refactoring can solve a non-existent problem.
More than that, CCP tends to increase with time \cite{Amit2021CCP, Amit2019Refactoring}, and therefore improvement and reduction-to-mean do not explain the common behavior.
Upon checking our recommendation by the CCP group, the function extractions are beneficial in all but the low CCP group and the `Suitable McCabe refactor' removal of too-many-return-statements is beneficial in the high group.
This is aligned with the old programming advice: ``If it is not broken, do not fix it''.

As another validation, we used another process metric of commit duration, measuring the time needed to modify the code \cite{Amit2021CCP, Amit2020Effort}.
It is reduced in the high and low groups but increases in the medium group, hence not monotonic with CCP in this context.
It is also reduced in all functions adding and suitable complexity reducing alerts, and showing benefits not related to the CCP metric.

\subsubsection{Mathematical Justification of Using the Hit Average}

Note that given a rule, we compute the average of the difference in all events identified by it.
Since our rules do not have perfect accuracy, this average could have been misleading since it includes misclassified alert removals.
Amit and Feitelson showed that using a confusion metric with the predictive performance of a classifier, one can find a maximum likelihood estimator of the positive rate using the hit rate \cite{Amit2021CCP}.
Although we could use a similar approach, our case is actually simpler.
First, note that only hits of the rule are being averaged.
Therefore, false negative mistakes do not influence the value of the average.
The recall of the rules, presented in Table \ref{tab:lf-pred-performance}, is high and protects us from using too few cases.
As for hits, we are averaging true positives and false positives instead of just true positives.
The true positives are the cases that should be averaged, and the high precision indicates that they are most of the cases (see Table \ref{tab:lf-pred-performance} again)  and therefore the influence of false positives is smaller.
More than that, Table \ref{tab:ccp-diff}, shows that control commits lead to a small increase in the tendency to bugs.
If the false positives behave like the control, averaging them with the true positives will lead to an upper bound on the average.
This is likely since the control represents most commits.
If the opposite happens and the false positives reduce the tendency to bugs more than the rule true positives, hence they are very effective and should be examined as interventions on their own.

%\idea{Do the same for adding alerts?}

\subsection{Exploration using Supervised Learning}

We used a dataset of 328 complexity-reducing alert removals with sufficient commits, a third of it was used as a test set.
The positive rate is 41\%, indicating that most interventions do not reduce the tendency to bugs.
The accuracy is important in supervised learning of changes, as if it is perfect, we are assured there are no other causal factors \cite{AMIT2025Survey}.
Our highest accuracy was 78\% by a high-capacity GradientBoostingClassifier, a good performance yet that indicates that there is still more to explore.
Logistic regression, a low capacity model, reached an accuracy of 75\%, indicating that the threat of overfitting is low.
When aiming for precision, a size-bounded Decision Tree reached precision of 82\% with recall of 47\%. 
Hence, it can be used to identify good interventions, while covering almost half of the good ones.
When targeting recall, a different Decision Tree reaches almost symmetric performance of 86\% recall and precision of 42\%.

\begin {table*}[h!]\centering
\caption{  \label{tab:features-cm} Features Predictive Performance }
\begin{tabular}{ | l| l| l| l| l| l  | }
\hline
Feature & Accuracy & Hit Rate & Precision & Precision Lift & Recall \\
\hline
mostly delete & 0.59 & 0.05 & 0.67 & 0.56 & 0.07\\ \hline
high ccp group & 0.63 & 0.25 & 0.61 & 0.43 & 0.36\\ \hline
low McCabe max before & 0.61 & 0.20 & 0.59 & 0.38 & 0.28\\ \hline
low McCabe sum before & 0.59 & 0.24 & 0.54 & 0.26 & 0.31\\ \hline
low McCabe max diff & 0.58 & 0.19 & 0.51 & 0.19 & 0.23\\ \hline
is refactor & 0.58 & 0.35 & 0.51 & 0.19 & 0.42\\ \hline
new function & 0.58 & 0.36 & 0.50 & 0.18 & 0.42\\ \hline
massive change & 0.57 & 0.16 & 0.50 & 0.17 & 0.19\\ \hline
only removal & 0.57 & 0.09 & 0.46 & 0.09 & 0.09\\ \hline
too-many-statements & 0.54 & 0.52 & 0.47 & 0.09 & 0.57\\ \hline
too-many-branches & 0.55 & 0.27 & 0.46 & 0.07 & 0.29\\ \hline
high McCabe sum diff & 0.54 & 0.20 & 0.42 & -0.02 & 0.20\\ \hline
low McCabe sum diff & 0.53 & 0.21 & 0.40 & -0.07 & 0.19\\ \hline
high McCabe sum before & 0.52 & 0.24 & 0.39 & -0.08 & 0.22\\ \hline
high McCabe max before & 0.52 & 0.22 & 0.39 & -0.09 & 0.20\\ \hline
too-many-nested-blocks & 0.55 & 0.09 & 0.37 & -0.14 & 0.08\\ \hline
high McCabe max diff & 0.56 & 0.03 & 0.30 & -0.30 & 0.02\\ \hline
low ccp group & 0.42 & 0.30 & 0.24 & -0.44 & 0.16\\ \hline
too-many-return-statements & 0.51 & 0.11 & 0.22 & -0.48 & 0.06\\ \hline
\end{tabular}
\end{table*}

Table \ref{tab:features-cm} presents the predictive power of various features of extraction alert removals, with the reduction of the tendency to bugs as the concept as before.
Unlike the models above, here we treat each feature as an independent classifier, ignoring the context.
The table allows us to understand the predictive performance of each feature, and its general contribution to the modeling.
The features are ordered by precision lift, the improvement in precision over the positive rate.

`mostly delete' has the highest precision and `only removal' has a positive lift too, though they usually do not represent a refactor.
We created `high' features, the top 25\% and `low', for the bottom 25\%.
`high ccp group' has the second highest lift.
Next, we have the low McCabe features. 
That might be surprising, contrary to the assumption that more complexity leads to more bugs.
A possible explanation comes from manual interventions.
Some files of high complexity are of hundreds or even more than 1,000 lines.
Perfecting a 50-lines-method has a small impact on the file since most commits are simply not related to it.
Hence, compared to the high complexity of the entire file, the complexity reduction is minor.
`is refactor', a refactor indication in the commit message, and
adding a new function, as used in our rules, both have the highest recall and high precision lift.
Massive change, which we filtered out in our rules, has a positive impact, hence this direction of future research might be beneficial.
`too-many-statements' and `too-many-branches' also have a positive lift.
The rest have a negative lift with `too-many-return-statement' being the lowest, one below `low ccp group'.

One can use our analysis to find new beneficial actions.
For example, combining useful features can lead to recommendations like intervene aiming to refactor (`is refactor') in error-prone files (`high ccp group') yet not too complex (`low McCabe sum before') using method extraction (`new function') to reduce the tendency to bugs.
Many of the models can lead to more actions, like by extracting paths in a decision tree.

\section{Threats}

\subsection{Manual Intervention Threats}

A single developer did the interventions in the current dataset.
Although we used Pylint to ensure that the alerts do not depend on the developer, the intervention is.
In many cases, the IDE provides a way to remove the alert (e.g., method extraction utility), indicating the possibility of a fixed mechanized solution.
Still, there is personal judgment (e.g., deciding which content to extract as a method).
%Even a basic case of a SQL query in line-too-long can be modified into just two lines or multiple lines that respect the query structure. 
%More developers are needed to identify personal habits and find out which ones are important. 
As a partial way to cope with this threat, we used Copilot with both ``Claude 3.5 sonnet'' and ``GPT 4o'' in a few repositories.
This added different approaches.
%Yet, the LLMs' output depends on the prompt, might differ in activation, or might not solve the alert correctly. 

Note that the use case of static analyzers assumes that one should fix the alerts, regardless of one's skill, the project profile, or the containing file content.
%The use of natural alert removals, from plenty of projects by plenty of developers, also copes with this threat.
In principle, our automated dataset could be used to evaluate our assumption.
However, there are a few repositories or developers that fixed the same alert a few times in our dataset.
There are 5 out of 6 (83\%) repositories in which too-many-branches were removed with function extraction at least 3 times in which the CCP was reduced, giving some support that the effect is not repository-specific but more general.

\hide{Like many other datasets in software engineering, our dataset violates the IID assumption (Independent
and Identically Distributed random variables), critical for statistical analysis.
For example, developers tend to follow specific patterns, and we have encountered cases of fixing an alert in almost similar code a few times in the same repository.
}

Some of the alerts are correlated.
When extracting a method to fix too-many-branches one might also fix too-many-nested-blocks since they are due to the same code.
This requires being careful with the attribution of changes in process metrics.

Static analysis treats alerts as binary.
However, removing 3 extra branches is not equivalent to removing 15.
This threat is reduced by considering the McCabe metric.

\subsection{Intervention-Like Events Threats} 

The method we present is new, so there is no direct prior work that can be given to support it.
However, the use of labeling functions has a solid mathematical background and successful applications \cite{schapire1990strength, Blum:1998:CLU:279943.279962, 10.5555/3157382.3157497, Amit2017Archimedes, Amit2024Motivation}.

We built our labeling functions using domain knowledge and hard-to-reproduce steps for one not familiar with the method.
Since labeling functions need to be only better than random, many times a weak signal (e.g., interventions vs. other commits) is enough to train classifiers that will find such functions.
In our case, one can build a model that distinguishes between manual interventions and general commits.
An interpretable model such as a Decision Tree \cite{quinlan2014c4} can lead to potential functions.
However, once found and validated, the source of the labeling functions is not important.

\hide{
Our labeling functions are better than random, but not perfectly accurate.
There are false-positive events mistakenly considered similar to an intervention.
When we compute the CCP difference, we include these functions' false positives too.
Since the tendency to bugs usually increases over time \cite{Amit2021CCP, Amit2019Refactoring}, while our estimate might be inaccurate, in this case, the true impact will be even higher.
}

Many intervention experiments compare the states before and after the intervention, assuming that there are no other influencing events.
We noted cases where alerts were added and removed a few times.
Only 70\% of the cases had at most two commits that changed an alert in a given file.
2.3\% even had 10 such commits or more.
That might happen when a function length is about the alert threshold and it is being modified, without the intention to refactor, so its length increases and decreases.
%Our use of the labeling functions on all commits enables identifying these cases and deciding how to treat them.

\subsection{General Threats}

The decision of a developer to remove an alert depends on many factors.
After selecting alerts for manual intervention using the uniform distribution, the owner notified us that we sometimes intervened in obsolete code that was not intended for modification.
This is a kind of domain adaptation problem \cite{ben2006analysis}, a common threat in causality \cite{chernozhukov2024applied}.
In this case, it means that the uniform distribution that we used does not represent well the selection of alerts to remove.
Those interested in such an analysis can model the alert selection and use the model for intervention selection.

In general, when natural events are used, their distribution is the distribution of interest, and domain adaptation does not occur.
However, biases of the labeling functions might lead to domain adaptation (e.g., not representing very large commits).
However, the usage of static analysis assumes that removing alerts is beneficial, regardless of context.

The number of cases that we considered is large with respect to intervention experiments, but is still moderate.
We deliberately used a simple analysis that has a low VC dimension \cite{vapnik1971VC}, reducing the threat of misalignment between the dataset and the data source.
Note that we used only Python files that were modified between two specific dates.
There is a wealth of more data that can be utilized, leading to much larger datasets.

When we considered the tendency to bugs, we divided the timeline into before and after the intervention.
It is likely that an improvement a year after the intervention is not due to it.
However, using a short duration leads to fewer commits, reducing robustness to noise.
Hence, it seems that there is not a single optimal duration window but windows that fit certain use cases. 
We checked the influence of removing alerts on the tendency to bugs.
If we required at least 10 commits in 3 months (e.g., as on the 749,603 commits in \cite{Amit2019Refactoring}), we would have get only 1,344 samples, trading noise per sample in noise due to dataset size.

When analyzing impact, one should consider indirect influence.
During our manual intervention, while extracting a method, we made the CI/CD identify an unused variable (in a new location), which was fixed.
We used negative control alerts to reduce this threat.
Removal of certain alerts might have a low benefit that is hard to notice or a benefit in other aspects.

Last, we applied supervised learning to the alert removal dataset, which includes non-refactoring actions.
For example, the ``only removal'' commits cannot remove an alert and be a refactoring, unless they remove unused code.
If new rules are extracted from supervised learning, they should be validated as we did for the rules presented in Table \ref{tab:lf-pred-performance}.

\hide{
\section{Future Work}

The programming language-specific work required for this research (e.g., wrapping the static analyzer) is significant.
We therefore used only Python code, leaving the question of generalization to other programming languages unanswered.
Note that our alerts of interest, from line-too-long to function complexity, appear in other programming languages.
Hence, performing the same analysis on repositories of a different language can provide external validation. 

The ability to evaluate machine learning models using real-world data allows analyzing their performance and further improving them.
Once we have a large dataset of interventions and their outcomes, we can apply causality classifiers and even regular machine learning classifiers trained to predict impact (e.g., \cite{AMIT2025Survey}).
Such classifiers make assumptions (e.g., the Markov property \cite{dynkin1965markov}), which might not fit the reality.
Evaluating them on data will allow knowing if they are part of the George Box ``wrong but useful models'' \cite{BOX1979201}.
Having a benchmark dataset for causality is considered to be an important challenge \cite{cinelli2025challengesstatisticsdozenchallenges}.

We observe a code change, on which developers work, and a change in the tendency to bugs.
We can use our intervention cases for direct comprehension experiments \cite{feitelson2022considerations, siegmund2016program}.
That allows comparing correctness and comprehension time with and without alert (e.g., unneeded parentheses).
Such experiments allow us to add in-vitro experiments (in glass, like using the alert in comprehension experiments) to our in-vivo experiments(in life, as in live GitHub repositories). 
This is a direct evaluation of a link in the causal chain.
Not only we will show a reduction in the McCabe complexity and a reduction in the tendency to bugs.
The in-vitro experiments can figure out if developer error more and need more time to comprehend the pre-intervention versions, explaining the higher tendency to bug in the in-vivo analysis. 
}

\section{Conclusions}

We presented new methods that enable high-scale causality analysis based on interventions. 
The size of the dataset is a constraint in manual intervention research.
We overcame the constraint by profiling interventions and using events that are similar to them.
We demonstrated the method in the domain of software quality, investigating the impact of static alert removal on the tendency to bugs.

Code metrics capture the change in the source code, a clean state influenced only by the manually intervening developer.
We found that removing function complexity alerts reduces McCabe complexity by 5.6 to 13.6, equivalent to $2^{5.6}=48$ to $2^{13.6}=12,416$ different paths to test.
This agrees with best practices, quantitatively showing that the impact is high and clarifying the mechanism by which interventions can influence the behavior captured by process metrics.

Identifying alert removals in nature, we built a dataset more than 15 times larger than our dataset of manual interventions.
Error bounds such as the VC dimension bound \cite{vapnik1971VC}, confidence intervals,  and many others are $O(\frac{1}{\sqrt{n}})$, where $n$ is the number of samples.
That means that a 15-times larger dataset is translated into a about 4-times tighter error bound.
Looking at the Corrective Commit Probability, a process metric of the tendency to bugs, we identified complexity-reducing interventions that reduce them significantly.
In 33\% of the February files analyzed, there was at least one complexity alert.
These are simple, actionable, and automatically identifiable recommendations that programmers can use to improve their code by fixing common problems.
The high effort needed for manual interventions is not unique to software engineering.
Our method allows us to profile interventions, identify similar events in nature, evaluate the profiling performance and improve it, and lastly analyze high-volume intervention datasets. 
If a domain has causality questions, many digitally documented events, and candidate interventions, our method should be valuable for it.
Examples are medicine (medication prescription), economics (new tax), and social sciences (new digital activity).
All of these will benefit from identifying causal relations by investing a low manual effort while obtaining robust results.

\section{Data Availability}

All data and code are provided in our repository \cite{Amit2025Pylint}.
We encourage researchers to further increase our dataset of manual interventions by following our intervention protocol and submitting pull requests with their interventions.

%A frozen version is available at \url{https://zenodo.org/records/17273031}

\bibliography{abbrv.bib, bibtex.bib}
\bibliographystyle{IEEEtran}

\end{document}